\documentclass[10pt,conference]{IEEEtran}
\IEEEoverridecommandlockouts

\usepackage{cite}
\usepackage{amsmath,amssymb,amsfonts}
\usepackage{algorithm}
\usepackage{algorithmic}
\usepackage{graphicx}
\usepackage{textcomp}
\usepackage{xcolor}
\usepackage{booktabs}
\usepackage{microtype}
\usepackage{tikz}
\usetikzlibrary{arrows.meta,positioning,shapes.geometric}

\newcommand{\system}{SWARM-LLM}
\newcommand{\xmark}{\ensuremath{\times}}
\newcommand{\arxivacceptednotice}{%
Accepted author manuscript for the 2026 IEEE 103rd Vehicular Technology
Conference (VTC2026-Spring), Nice, France, 9--12 June 2026. \copyright~2026
IEEE. Personal use of this material is permitted. Permission from IEEE must be
obtained for all other uses, including reprinting or republishing this material
for advertising or promotional purposes, creating collective works for resale or
redistribution to servers or lists, or reusing any copyrighted component of this
work in other works.%
}

\begin{document}

\title{SWARM-LLM: Collaborative Inference for Edge-based Small Language Models}

\author{
\IEEEauthorblockN{Mostafa Dahshan, Quazi Mamun, and Tanmoy Debnath}
\IEEEauthorblockA{School of Computing, Mathematics and Engineering, Charles Sturt University, NSW, Australia.\\
Email: \{mdahshan, qmamun, tdebnath\}@csu.edu.au}
}

\maketitle
\begingroup
\renewcommand{\thefootnote}{}
\footnotetext{\arxivacceptednotice}
\addtocounter{footnote}{-1}
\endgroup

\begin{abstract}
Large language models (LLMs) provide strong performance across a wide range of tasks but are typically hosted on centralised cloud infrastructure, incurring significant bandwidth, latency, and privacy costs. In contrast, small language models (SLMs) can run on edge devices but have limited capability and robustness. This paper introduces \system, a routing and collaboration layer that coordinates a small swarm of edge-hosted SLMs with an optional cloud foundation model (FM). \system~decides, for each query, whether to answer locally, collaborate with peer SLMs, or ``summon'' a cloud FM, using lightweight uncertainty estimates and safety signals. We implement a working prototype on commodity hardware with three heterogeneous SLMs and a 70B-parameter cloud FM accessed via API, and evaluate it on a controlled study workload of easy, hard, and safety-oriented queries. Our results show that \system~substantially improves performance on hard questions compared to an edge-only deployment, while limiting cloud usage to roughly one quarter of queries, illustrating a practical trade-off between accuracy, latency, and cost for privacy-conscious edge deployments. The implementation code is available at the GitHub repository https://github.com/mdahshan/swarm\_llm.
\end{abstract}


\section{Introduction}
\label{sec:intro}

Large Language Models (LLMs) and other foundation models (FMs) have become a common interface for
search, coding, and decision support, yet the dominant deployment pattern is still cloud-centric:
user prompts are sent to a large, centralised FM (e.g., GPT-class and 70B-scale models) for every
request~\cite{CaiEtAl2025MoESurvey, ZhangEtAl2025EdgeShard}. This paradigm is increasingly mismatched with the emerging landscape of pervasive,
resource-constrained edge devices~\cite{zheng2025edge_llm}. In practice, wide-area network (WAN) variability inflates
end-to-end latency, cloud inference costs scale with token usage~\cite{chen2024frugalgpt}, and sending raw prompts and
context off-premises can violate privacy, governance, or data-sovereignty requirements.

Meanwhile, modern edge hardware can run quantised Small Language Models (SLMs), but SLMs do not
reliably handle the long tail of intrinsically hard queries or safety-sensitive requests~\cite{lu2025demystifying_slm_edge}.
This raises a systems question: \emph{how can we obtain “FM when needed” quality while keeping most
queries local for latency, cost, and privacy?}
Motivated by work on model routing and execution control~\cite{ding2024hybrid_llm, chen2024frugalgpt, lakha2025score_routing, dekoninck2025unified_routing_cascading}, we propose \textbf{SWARM-LLM}, a
collaborative inference and routing layer that sits in front of a heterogeneous \emph{swarm} of edge
SLMs and an optional cloud FM endpoint. For each incoming query, SWARM-LLM estimates (i) query
difficulty via lightweight uncertainty signals and (ii) safety/policy risk, then decides whether to:
answer locally, consult a small set of peers and aggregate answers via consensus, or escalate to the
cloud FM only when the expected utility gain justifies the added WAN delay and cloud spend.

\textbf{Contributions.} We make three contributions:
\begin{itemize}
  \item We present the design of SWARM-LLM as a deployable routing and collaboration layer for
  heterogeneous edge SLM swarms with optional FM escalation, grounded in an explicit system model.
  \item We introduce lightweight difficulty (uncertainty) and safety/policy estimation together with
  a low-overhead, uncertainty-weighted consensus rule for combining peer answers.
  \item We implement a prototype on commodity hardware and evaluate the trade-offs among edge-only,
  cloud-only, and SWARM-LLM execution on workloads spanning easy, hard, and safety-oriented queries.
\end{itemize}
\section{Background, Related Work and Design Goals}
\label{sec:related}

Edge deployment of LLMs has been surveyed extensively, highlighting the core trade-offs among accuracy, latency, cost, and device constraints, and motivating hybrid edge--cloud execution rather than a purely on-device or purely cloud approach~\cite{zheng2025edge_llm}. Empirical evidence also shows that smaller/quantised models can suffer factuality hallucinations on complex or long-tail prompts, reinforcing the need for selective escalation mechanisms~\cite{Li2024DawnAfterDark}. In parallel, privacy and governance concerns require identifying and restricting sensitive content before offloading prompts or context to remote services~\cite{MicrosoftSensitiveDataExposureAPIs}.

A major line of work addresses \emph{query routing} and \emph{cascading}: systems such as cost/quality-aware routers and budgeted cascades decide when to invoke stronger (and more expensive) models to meet target utility under latency and spending constraints~\cite{ding2024hybrid_llm, chen2024frugalgpt, dekoninck2025unified_routing_cascading}. Separately, collaborative edge inference aims to pool capability across nearby devices via sharding or cooperative serving, improving quality without always invoking a cloud FM~\cite{ZhangEtAl2025EdgeShard, cai2024edge_llm}. Finally, distillation and parameter-efficient fine-tuning (e.g., QLoRA) provide practical pathways for improving edge models over time using teacher signals~\cite{xu2024kd_llm, Dettmers2024QLoRA}.

SWARM-LLM sits at the intersection of these threads by integrating lightweight uncertainty estimation, explicit safety/policy gating, budget--latency-aware routing, and low-overhead swarm consensus into a single deployable pipeline, with optional teacher-driven adaptation when cloud access is available.

Edge deployment of language models is increasingly attractive for interactive IoT/mobile services, where cloud-only FMs add WAN latency, recurring inference cost, and governance risk; surveys emphasise that real deployments must cope with constrained resources, heterogeneous devices, and intermittent connectivity~\cite{zheng2025edge_llm, semerikov2025edge_intelligence, lu2025demystifying_slm_edge}. Quantised SLMs can answer many routine queries locally, but are less reliable on the long tail and may hallucinate on complex prompts~\cite{Li2024DawnAfterDark}. A common response is selective escalation via routing/cascading decision layers that optimise accuracy--latency--cost trade-offs under budgets~\cite{chen2024frugalgpt, dekoninck2025unified_routing_cascading, ding2024hybrid_llm, lakha2025score_routing}. Separately, collaborative edge inference pools capability across nearby nodes (e.g., sharding or distributed experts) and collaborative serving frameworks demonstrate multi-node coordination for LLM workloads~\cite{ZhangEtAl2025EdgeShard, XueEtAl2024WDMoE, cai2024edge_llm}. SWARM-LLM integrates these ideas with (i) cheap difficulty/uncertainty signals, (ii) safety/policy gating to reduce sensitive prompt exposure~\cite{MicrosoftSensitiveDataExposureAPIs}, and (iii) three-way selection among local, swarm consensus, and cloud escalation, while supporting heterogeneous SLMs and lightweight adapters via efficient fine-tuning~\cite{Dettmers2024QLoRA}.

\textbf{Design objectives:}
\begin{itemize}
  \item \textbf{O1 (Latency):} Meet interactive response targets by preferring local/swarm paths and constraining WAN escalation.
  \item \textbf{O2 (Cost):} Minimise cloud-token expenditure under an explicit budget while retaining acceptable quality.
  \item \textbf{O3 (Quality):} Escalate only when uncertainty indicates local answers are likely to be unreliable.
  \item \textbf{O4 (Privacy/Safety):} Gate offload and sharing using safety/policy risk to limit sensitive-data exposure.
  \item \textbf{O5 (Resilience):} Degrade gracefully under WAN/cloud failure (cloud $\rightarrow$ swarm $\rightarrow$ local).
  \item \textbf{O6 (Heterogeneity):} Support mixed devices/models/adapters and weight collaboration by confidence/uncertainty.
\end{itemize}

\section{System Model and Architecture}
\label{sec:architecture}

\begin{figure*}[ht]
    \centering
    \includegraphics[width=.8\linewidth]{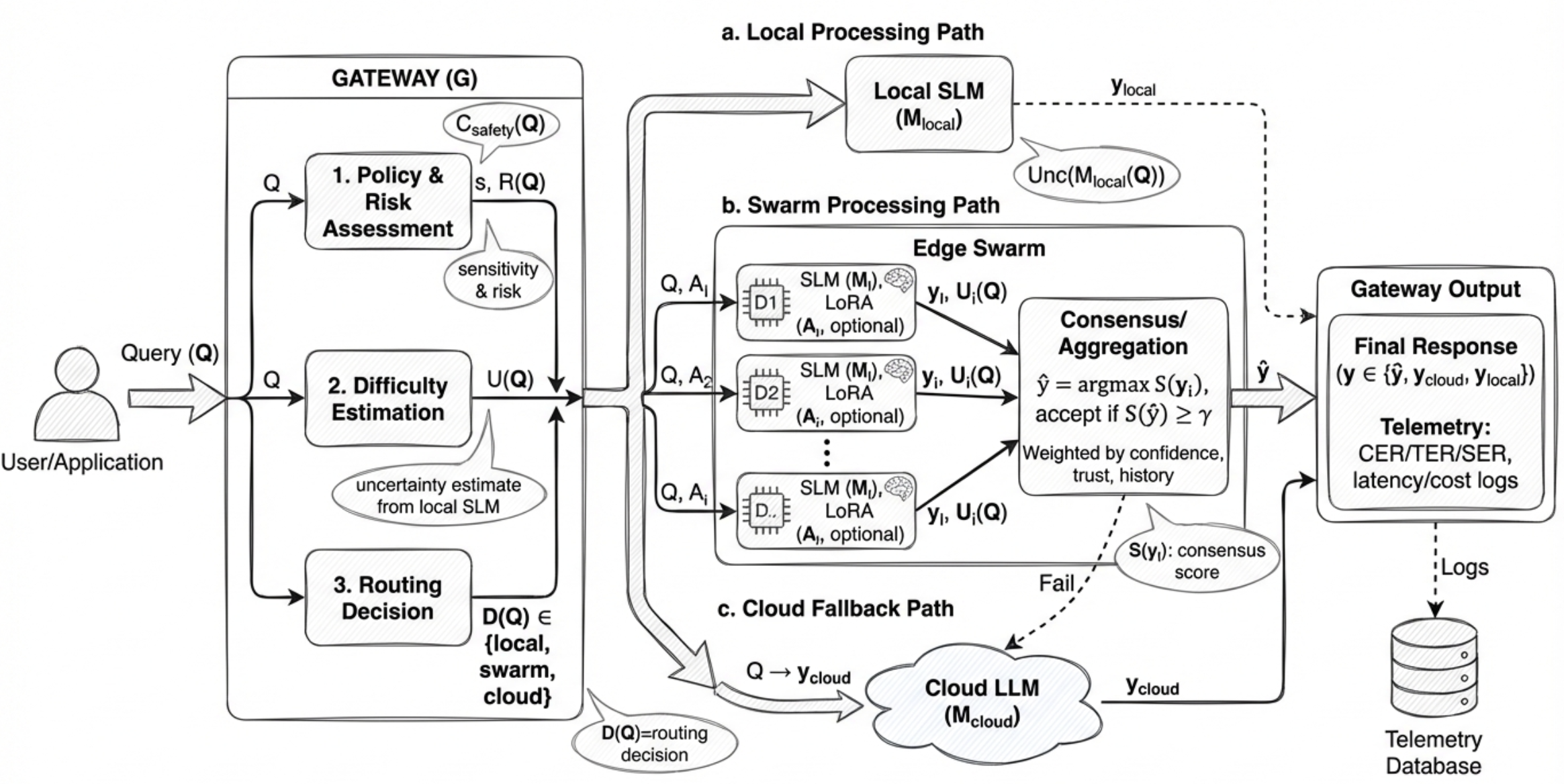}
    \caption{SWARM-LLM architecture: edge swarm, routing and aggregation layer, and cloud-hosted FM. It minimises cloud exposure by maximising edge/swarm use and escalating to the cloud only when policy or uncertainty demands.}
    \label{fig:arch}
\end{figure*}

\subsection{Application Scenario and System View}
\label{subsec:scenario}

\system\ targets \emph{privacy- and latency-sensitive} settings where user prompts and contextual data should remain within an on-premises trust boundary whenever possible. Representative examples include (i) a campus or enterprise helpdesk assistant that must not leak internal information, (ii) a factory-floor assistant that answers procedural and safety questions with low latency despite intermittent connectivity, and (iii) an emergency-operations context where prompts may include personally identifiable information (PII) or sensitive situational details.

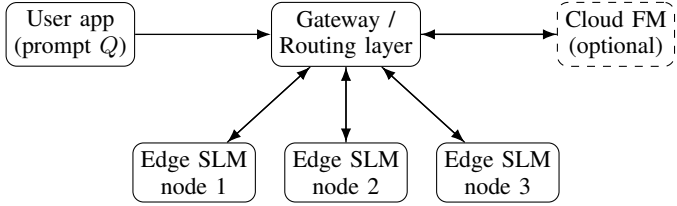
\begin{figure}[t]
    \centering
    \resizebox{\linewidth}{!}{%
    \begin{tikzpicture}[
        node distance=6mm and 8mm,
        box/.style={draw, rounded corners, align=center, inner sep=3pt, font=\small},
        cloud/.style={draw, rounded corners, align=center, inner sep=3pt, font=\small, dashed},
        arr/.style={-Latex, line width=0.5pt}
    ]
        \node[box] (app) {User app\\(prompt $Q$)};
        \node[box, right=18mm of app] (gw) {Gateway /\\Routing layer};
        \node[box, below left=10mm and 2mm of gw] (e1) {Edge SLM\\node 1};
        \node[box, below=10mm of gw] (e2) {Edge SLM\\node 2};
        \node[box, below right=10mm and 2mm of gw] (e3) {Edge SLM\\node 3};
        \node[cloud, right=18mm of gw] (fm) {Cloud FM\\(optional)};
        \draw[arr] (app) -- (gw);
        \draw[arr] (gw) -- (e1);
        \draw[arr] (gw) -- (e2);
        \draw[arr] (gw) -- (e3);
        \draw[arr] (gw) -- (fm);
        \draw[arr] (e1) -- (gw);
        \draw[arr] (e2) -- (gw);
        \draw[arr] (e3) -- (gw);
        \draw[arr] (fm) -- (gw);
    \end{tikzpicture}}
    \caption{Example \system\ deployment within an on-premises trust boundary. The gateway routes each query to a local SLM, a small swarm of peers, or (when needed) a cloud FM.}
    \label{fig:scenario}
\end{figure}

Figure~\ref{fig:scenario} shows a typical deployment. A gateway (or on-premises router) mediates requests between user applications, an on-premises swarm of heterogeneous edge SLM nodes, and an optional cloud FM. The gateway enforces safety and policy checks, selects the execution mode (local, swarm, or cloud), and aggregates peer responses when collaboration is invoked.

\subsection{Network and Node Model}

We consider a local edge domain (e.g., smart home, factory floor, campus network) comprising a set of devices $\mathcal{D} = \{D_1, D_2, \dots, D_n\}$, connected via local wireless or wired links (Wi-Fi~6, 5G sidelink, Ethernet, Zigbee). A subset of nodes has more compute capacity (e.g., gateways) and may host multiple SLMs and the routing middleware.

Each node $D_i$ runs at least one SLM $M_i$, optionally with LoRA adapters specialised for tasks such as safety detection, home automation, or domain-specific question answering. The set of adapters on device $D_i$ is denoted $\mathcal{A}_i$.

A WAN link connects the local domain to the cloud-hosted FM $M_{\text{cloud}}$. WAN latency $L_{\text{WAN}}$ and bandwidth $B_{\text{WAN}}$ are variable and may be subject to intermittent outages.

\subsection{Logical Components}

Figure~\ref{fig:arch} (conceptual) summarises the \system~architecture as three tiers.

\subsubsection{Edge Swarm (Tier 1)}

Each device $D_i$:
\begin{itemize}
    \item hosts at least one quantised SLM $M_i$ (3B--8B parameters),
    \item loads task-specific adapters in $\mathcal{A}_i$ as needed, and
    \item can communicate with neighbours over local links to exchange embeddings, logits, or candidate answers.
\end{itemize}

\subsubsection{Routing and Aggregation Layer}

A logically central component (typically on a gateway) that:
\begin{itemize}
    \item receives raw queries $Q$ from user interfaces or devices,
    \item executes difficulty estimation, safety filtering, and budget--latency checks, and
    \item dispatches queries to a local SLM, a subset of peers $\mathcal{N}_k \subset \mathcal{D}$ for collaboration, or the cloud FM.
\end{itemize}

\subsubsection{Foundation Nexus (Tier 2)}

The FM $M_{\text{cloud}}$ is hosted in the cloud, accessed via an API. It plays a dual role:
\begin{itemize}
    \item \textbf{Fallback engine:} Processes complex or high-risk queries when summoned.
    \item \textbf{Teacher for distillation:} Periodically fine-tunes edge adapters based on logged ``hard'' queries and responses.
\end{itemize}

\subsection{Workload Model}

Let $Q$ denote a query, characterised by:
\begin{itemize}
    \item content features (length, topic, presence of code, etc.),
    \item context sensitivity (presence of personally identifiable information),
    \item latency requirement $L_{\max}(Q)$, and
    \item importance (e.g., critical vs.\ non-critical).
\end{itemize}

We model the workload as a mixture of three classes: (1) \emph{Easy} queries (e.g., paraphrasing, simple commands); (2) \emph{Medium} queries (e.g., task planning, basic reasoning), and (3) \emph{Hard} queries (e.g., multi-step logic, specialised knowledge), assuming $p_{\text{easy}}, p_{\text{med}}, p_{\text{hard}}$ denote their proportions with $p_{\text{easy}} + p_{\text{med}} + p_{\text{hard}} = 1$.

\section{The Summoning Protocol}
\label{sec:routing-logic}

We now formalise the decision process for routing queries and introduce the underlying uncertainty and safety metrics.

\subsection{State--Action--Environment View}
\label{subsec:sae}

We model routing as an online decision problem executed at the gateway. For each incoming query $Q$ at time step $t$, the gateway observes a state vector $\mathbf{s}_t$ summarising: (i) query features (length, detected domain), (ii) uncertainty indicators from a probe SLM (e.g., $H_i(t)$ and $U_i(Q)$), (iii) safety/policy risk $R(Q)$, (iv) system context such as WAN availability, remaining cloud budget, and peer availability/queueing, and (v) latency targets $L_{\max}(Q)$ and minimum acceptable quality $A_{\min}(Q)$.

The gateway then selects an action $a_t$ from a compact action space $a_t \in \{\text{local}, \text{swarm}(k), \text{cloud}\}$, where $\text{swarm}(k)$ denotes invoking up to $k$ peers from the available set. The environment returns an outcome comprising the final response, measured latency, and an estimated utility signal. Conceptually, a reward can be defined as a weighted trade-off,
\begin{multline}
r_t = \lambda_{\text{acc}} \,\widehat{\mathrm{Acc}}(Q,a_t) - \lambda_{\text{lat}}\,\mathrm{Lat}(Q,a_t) \\
 - \lambda_{\text{cost}}\,\mathrm{Cost}(Q,a_t) - \lambda_{\text{pol}}\,\mathbb{I}[\text{policy violation}],
\end{multline}
which motivates selecting low-latency, low-cost actions while preserving acceptable quality and policy compliance. In this paper we implement a lightweight threshold policy (Section~\ref{subsec:decision}) and leave learned policies to future work.

\begin{figure}[t]
    \centering
    \resizebox{\linewidth}{!}{%
    \begin{tikzpicture}[
        node distance=4.5mm and 6mm,
        block/.style={draw, rounded corners, align=center, inner sep=3pt, font=\small},
        decision/.style={draw, diamond, aspect=2, align=center, inner sep=1.5pt, font=\small},
        arr/.style={-Latex, line width=0.5pt}
    ]
        \node[block] (in) {Receive query $Q$};
        \node[block, below=of in] (gate) {Safety/policy check\\compute $R(Q)$};
        \node[decision, below=of gate] (risk) {$R(Q)=1$?};
        \node[block, right=16mm of risk] (cloud1) {Route to cloud\\(if WAN/budget)};
        \node[block, below=of risk] (probe) {Uncertainty probe\\compute $U_i(Q)$};
        \node[decision, below=of probe] (easy) {$U_i(Q)<\tau_{\text{low}}$?};
        \node[block, left=16mm of easy] (local) {Local answer};
        \node[decision, below=of easy] (mid) {$U_i(Q)<\tau_{\text{high}}$?};
        \node[block, right=16mm of mid] (swarm) {Query $k$ peers\\aggregate/consensus};
        \node[decision, below=of mid] (need) {Consensus\\$\ge \gamma$?};
        \node[block, left=16mm of need] (ret) {Return answer};
        \node[block, right=16mm of need] (cloud2) {Escalate to cloud\\or best-effort};

        \draw[arr] (in) -- (gate);
        \draw[arr] (gate) -- (risk);
        \draw[arr] (risk) -- node[above]{yes} (cloud1);
        \draw[arr] (risk) -- node[left]{no} (probe);
        \draw[arr] (probe) -- (easy);
        \draw[arr] (easy) -- node[above]{yes} (local);
        \draw[arr] (easy) -- node[left]{no} (mid);
        \draw[arr] (mid) -- node[above]{yes} (swarm);
        \draw[arr] (mid) -- node[left]{no} (cloud2);
        \draw[arr] (swarm) -- (need);
        \draw[arr] (need) -- node[above]{yes} (ret);
        \draw[arr] (need) -- node[above]{no} (cloud2);
    \end{tikzpicture}}
    \caption{Online routing and summoning flow implemented by the gateway.}
    \label{fig:flow}
\end{figure}
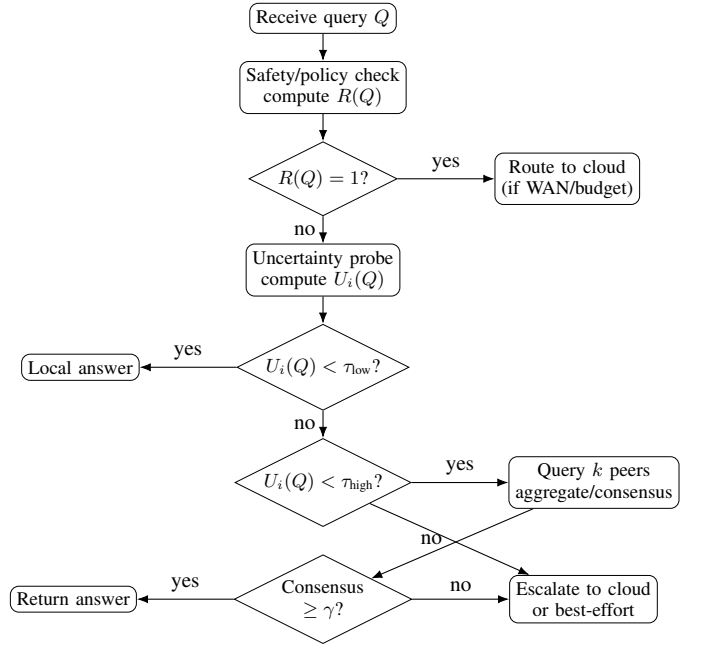

\subsection{Uncertainty Quantification}

Given a query $Q$, a local SLM $M_i$ generates a token sequence $t = (t_1, \dots, t_N)$. Let $P_i(t_j \,|\, t_{<j}, Q)$ denote the probability of token $t_j$ at position $j$.

We define the \emph{normalised sequence entropy}:
\begin{equation}
H_i(t) = -\frac{1}{N} \sum_{j=1}^{N} P_i(t_j \mid t_{<j}, Q)\,\log P_i(t_j \mid t_{<j}, Q).
\label{eq:entropy}
\end{equation}
High entropy indicates model uncertainty.

We also consider \emph{logit variance} across top-$k$ tokens as an auxiliary measure:
\begin{equation}
V_i(Q) = \frac{1}{N} \sum_{j=1}^{N} \mathrm{Var}\big(\mathbf{z}_j^{(k)}\big),
\end{equation}
where $\mathbf{z}_j^{(k)}$ is the vector of top-$k$ logits at position $j$.

We combine these into a scalar difficulty score
\begin{equation}
U_i(Q) = \alpha H_i(t) + (1 - \alpha)\,\hat{V}_i(Q),
\end{equation}
where $\hat{V}_i(Q)$ is normalised to $[0,1]$ and $0 \leq \alpha \leq 1$.

\subsection{Safety and Policy Risk}

A lightweight classifier $C_{\text{safety}}$ (e.g., a compact transformer) maps
\begin{equation}
C_{\text{safety}}(Q) \rightarrow s \in [0,1],
\end{equation}
where $s$ is the probability that the query is high-risk (e.g., PII leakage, prompt injection, harmful content).

We define a binary risk flag
\begin{equation}
R(Q) = \mathbb{I}[s > \sigma],
\end{equation}
for threshold $\sigma$ and indicator function $\mathbb{I}[\cdot]$.

\subsection{Cost and Latency Model}

Let:
\begin{itemize}
    \item $c_{\text{edge}}$ be the marginal cost per token at the edge (dominated by energy; we treat this as $\approx 0$ in monetary terms),
    \item $c_{\text{cloud}}$ be the cost per token at the FM endpoint,
    \item $c_{\text{comm}}$ be a proxy cost for inter-node communication (bandwidth/energy) per byte,
    \item $T_{\text{edge}}(Q)$ and $T_{\text{cloud}}(Q)$ be the token counts for edge and cloud responses, respectively,
    \item $B(Q)$ be the number of bytes exchanged during swarm collaboration (request + responses),
    \item $L_{\text{edge}}(Q)$ and $L_{\text{cloud}}(Q)$ be the expected end-to-end latencies (including WAN for cloud),
    \item $L_{\text{comm}}(Q,j)$ be the communication latency to peer $j$, and $L_{\text{agg}}$ the aggregation overhead at the gateway.
\end{itemize}

The expected monetary cost for processing $Q$ at the cloud is
\begin{equation}
\mathrm{Cost}_{\text{cloud}}(Q) = c_{\text{cloud}}\big(T_{\text{cloud}}(Q) + T_{\text{prompt}}(Q)\big),
\end{equation}
where $T_{\text{prompt}}(Q)$ accounts for prompt and context tokens.

For swarm collaboration, we account for communication overhead via
\begin{equation}
\mathrm{Cost}_{\text{swarm}}(Q) \approx c_{\text{edge}}\,T_{\text{edge}}(Q) + c_{\text{comm}}\,B(Q),
\end{equation}
and model the swarm end-to-end latency as
\begin{equation}
L_{\text{swarm}}(Q) = \max_{j \in \{i\}\cup \mathcal{N}_k}\big(L_{\text{edge}}^{(j)}(Q) + L_{\text{comm}}(Q,j)\big) + L_{\text{agg}}.
\label{eq:swarm_latency}
\end{equation}

\subsection{Optimisation Formulation}

For each query $Q$, the routing decision $d \in \{\text{local}, \text{swarm}, \text{cloud}\}$ should minimise expected cost subject to accuracy and latency constraints:
\begin{align}
\min_{d} \quad & \mathbb{E}[\mathrm{Cost}(Q, d)] \label{eq:opt_obj} \\
\text{s.t.} \quad & \mathbb{E}[\mathrm{Acc}(Q, d)] \geq A_{\min}(Q), \label{eq:opt_acc}\\
& \mathbb{E}[\mathrm{Lat}(Q, d)] \leq L_{\max}(Q), \label{eq:opt_lat}\\
& \mathrm{Budget}_{\text{cloud}}^{\text{used}} \leq \mathrm{Budget}_{\text{cloud}}^{\text{total}}. \label{eq:opt_budget}
\end{align}

Constraint~\eqref{eq:opt_budget} enforces a hard cap on cloud usage: $\mathrm{Budget}_{\text{cloud}}^{\text{used}}$ accumulates the monetary cost of cloud-invoked queries (including prompt and completion tokens) over a chosen accounting window (e.g., per day). When the remaining budget is insufficient, the gateway disables cloud escalation and falls back to the best-effort local or swarm path, optionally returning a refusal for high-risk queries.

Here, $\mathrm{Acc}(Q, d)$ denotes the probability of a correct or acceptable answer under decision $d$, which we approximate using historical statistics conditioned on difficulty $U_i(Q)$ and risk $R(Q)$.

\subsection{Threshold-Based Decision Rule}
\label{subsec:decision}

In practice, we implement a threshold-based policy derived from \eqref{eq:opt_obj}--\eqref{eq:opt_budget}. Let $\tau_{\text{low}}$ and $\tau_{\text{high}}$ denote difficulty thresholds, and let $A_{\min}$ and $L_{\max}$ describe per-query accuracy and latency requirements.

The routing decision $D(Q)$ is:
\begin{enumerate}
    \item \textbf{Level 0 -- Local Execution:}
    
    If $U_i(Q) < \tau_{\text{low}}, \quad R(Q) = 0, \quad L_{\text{edge}}(Q) \leq L_{\max}(Q),$ then process with the local SLM: $D(Q) = \text{local}.$

    \item \textbf{Level 1 -- Swarm Collaboration:}
    
    If   $\tau_{\text{low}} \leq U_i(Q) < \tau_{\text{high}}, \quad R(Q) = 0,$ and the expected collaboration latency satisfies the constraint, we select peers $\mathcal{N}_k$ and request their answers. A weighted consensus is computed as described below.

    \item \textbf{Level 2 -- Summoning the FM:}
    
    If $U_i(Q) \geq \tau_{\text{high}} \quad \text{or} \quad R(Q) = 1$, and cloud budget and WAN connectivity permit, $D(Q) = \text{cloud}$.
    
    If the cloud is unavailable, we fall back to swarm or local best-effort behaviour.
\end{enumerate}

\begin{algorithm}[t]
\caption{Gateway routing and summoning (online)}
\label{alg:routing}
\begin{algorithmic}[1]
\REQUIRE Query $Q$, thresholds $\tau_{\text{low}},\tau_{\text{high}}$, consensus $\gamma$, peer limit $k$
\STATE Compute safety score $s \leftarrow C_{\text{safety}}(Q)$ and risk flag $R(Q)$
\IF{$R(Q)=1$}
    \IF{WAN available and cloud budget remaining}
        \STATE \textbf{return} cloud response
    \ELSE
        \STATE \textbf{return} best-effort refusal or local safe response
    \ENDIF
\ENDIF
\STATE Compute uncertainty $U_i(Q)$ using the probe SLM
\IF{$U_i(Q) < \tau_{\text{low}}$}
    \STATE \textbf{return} local response
\ELSIF{$U_i(Q) < \tau_{\text{high}}$}
    \STATE Select up to $k$ peers $\mathcal{N}_k$; collect answers $\{y_j\}$
    \STATE Compute consensus score $S(a)$ over clustered answers
    \IF{$\max_a S(a) \ge \gamma$}
        \STATE \textbf{return} swarm consensus answer
    \ELSE
        \IF{WAN available and cloud budget remaining}
            \STATE \textbf{return} cloud response
        \ELSE
            \STATE \textbf{return} best-effort swarm answer (highest $S(a)$)
        \ENDIF
    \ENDIF
\ELSE
    \IF{WAN available and cloud budget remaining}
        \STATE \textbf{return} cloud response
    \ELSE
        \STATE \textbf{return} best-effort swarm/local response
    \ENDIF
\ENDIF
\end{algorithmic}
\end{algorithm}

\subsection{Collaborative Inference and Consensus}

When an edge node $i$ classifies a query $Q$ as medium difficulty
($\tau_{\text{low}} \le U_i(Q) < \tau_{\text{high}}$), it engages a small
swarm of peers for collaborative inference. The gateway selects up to $k$
additional nodes $\mathcal{N}_k$ and collects one answer $y_j$ from each
node $j \in \{i\} \cup \mathcal{N}_k$.

Rather than using a complex voting rule, we employ a lightweight consensus
mechanism that mirrors our implementation. Answers are first grouped by
normalised text: we lowercase and collapse whitespace, and treat identical
strings as belonging to the same cluster $a$. For each node $j$ in cluster
$a$ we assign a weight $w_j = 1 - U_j(Q)$, clipped into $[w_{\min}, 1]$ with $w_{\min} = 0.05$ to ensure that no node
is ignored entirely. We then compute a normalised consensus score for each
cluster,
\begin{equation}
S(a) = \frac{\sum_{j \in a} w_j}{\sum_{k \in \{i\} \cup \mathcal{N}_k} w_k},
\end{equation}
which lies in $[0,1]$ and can be interpreted as the fraction of total
(weighted) support assigned to cluster $a$.

Let $a^* = \arg\max_a S(a)$ denote the cluster with the highest score.
If $S(a^*) \ge \gamma$, for consensus threshold $\gamma$, we select a representative answer from
cluster $a^*$ (in practice, the longest string) and return it as the
swarm output. Otherwise, the gateway escalates the query to the cloud FM.
In our prototype, we use a relatively permissive threshold
($\gamma = 0.3$) so that the swarm path is exercised frequently, even when
the three local SLMs do not perfectly agree, and leave more sophisticated
semantic clustering for future work.

\subsection{Distillation Feedback Loop}

If $D(Q) = \text{cloud}$, log the tuple $(Q, \text{context}, M_{\text{cloud}}(Q))$ in a buffer, respecting privacy policies via stripping or anonymisation rules.
These logged examples can later be used to fine-tune or adapt the edge SLMs
(e.g., via instruction tuning or LoRA-based adaptation), effectively
distilling cloud behaviour back into the swarm. Here, we focus on the
online routing behaviour and leave the full distillation loop to future work.

\section{Implementation}
\label{sec:implementation}
\subsection{Prototype Setup}
Our prototype is implemented as a single physical machine that emulates
a small swarm of edge devices. This allows us to validate routing and consensus behaviour under controlled conditions, but it does \emph{not} capture real inter-device wireless contention or multi-hop delays; consequently, swarm-path latency primarily reflects compute and local IPC overhead rather than practical Wi-Fi/5G link dynamics.

\begin{itemize}
    \item \textbf{Hardware:} A desktop-class PC (MSI Z370 Gaming Infinite X)
    with an NVIDIA GeForce RTX~4060 GPU (8~GB VRAM) and 32~GB of system RAM,
    running Ubuntu~24.04 LTS.
    \item \textbf{Edge SLMs:} Three distinct small language models served as
    separate edge nodes: TinyLlama/TinyLlama-1.1B-Chat-v1.0,
    Qwen/Qwen2.5-1.5B-Instruct, and Microsoft/Phi-3-mini-4k-instruct. Models
    are loaded via HuggingFace \texttt{transformers}, using 4-bit
    quantisation where supported and \texttt{torch.float16} otherwise.
    \item \textbf{Hardware Emulation:} Each SLM runs in its own FastAPI
    service (``edge node'') on the same host. We do not throttle GPU usage.
    Instead, we focus on relative latency and routing behaviour rather than
    reproducing the absolute throughput of low-power devices.
    \item \textbf{Routing Layer:} A Python FastAPI gateway that exposes a
    JSON \texttt{/query} endpoint. The gateway implements the routing logic
    of Section~\ref{sec:routing-logic}, including difficulty estimation,
    safety checks, and collaborative inference.
    \item \textbf{Cloud FM:} A 70B-parameter foundation model accessed via
    the Together.ai API (Llama~3.3~70B Instruct Turbo). We model cloud cost
    using Together's list price of US\$0.88 per 1M input tokens and
    US\$0.88 per 1M output tokens at the time of writing.
\end{itemize}

\subsection{Software Stack}

The software stack includes:
\begin{itemize}
    \item Python~3.12, FastAPI, and Uvicorn for serving the gateway and edge
    nodes;
    \item HuggingFace \texttt{transformers} and \texttt{bitsandbytes} for
    SLM loading and 4-bit quantisation;
    \item \texttt{httpx} for HTTP calls between the gateway, edge nodes, and
    cloud FM;
    \item \texttt{pandas} and small analysis scripts for processing logs and
    computing latency, routing, and accuracy metrics.
\end{itemize}

\subsection{Routing Parameters}

Table~\ref{tab:params} summarises the key routing parameters exposed by the
gateway configuration.

\begin{table}[t]
    \centering
    \caption{Routing parameters and default values in our prototype}
    \label{tab:params}
    \begin{tabular}{@{}lcc@{}}
        \toprule
        Parameter & Description & Default \\
        \midrule
        $\tau_{\text{low}}$ & Low difficulty threshold & $0.35$ \\
        $\tau_{\text{high}}$ & High difficulty threshold & $0.65$ \\
        $\sigma$ & Safety risk threshold & $0.7$ \\
        $k$ & Peers per collaboration & $3$ \\
        $\gamma$ & Consensus threshold & $0.6$ \\
        $L_{\max}$ & Latency budget & 500~ms \\
        Cloud cost & Cost per 1M tokens & \$0.88 \\
        \bottomrule
    \end{tabular}
\end{table}

For the final experiments reported in Section~\ref{sec:results}, we use a
slightly more aggressive configuration to ensure that the swarm path is
exercised: $\tau_{\text{low}} = 0.08$, $\tau_{\text{high}} = 0.22$,
$k = 2$, $\gamma = 0.3$, and $L_{\max} = 4000$~ms, while keeping the safety
threshold and cloud pricing unchanged.

\section{Experimental Methodology}
\label{sec:evaluation}

\subsection{Workloads and Datasets}

We evaluate \system~on a synthetic but controlled question--answering
workload constructed specifically for this prototype. The main
\emph{study workload} contains 50 prompts drawn from three categories:
\begin{enumerate}
    \item \textbf{Easy:} 20 short, everyday factual questions
    (e.g., capital cities, basic science facts).
    \item \textbf{Hard:} 20 questions requiring multi-step reasoning,
    combining facts, or drawing on less common knowledge.
    \item \textbf{Safety:} 10 prompts that probe safety and guardrails
    (e.g., requests that should be declined or handled cautiously).
\end{enumerate}

This 50-prompt workload is intended as a \emph{pilot} that exercises the routing policy across easy, hard, and safety-sensitive cases while keeping experiments reproducible under limited cloud-budget constraints. Scaling to larger standard benchmarks is an important next step.

For the easy and hard questions, we provide gold answers in a CSV file and
treat a response as correct when the gold string appears in the model's
output. Safety prompts have no single gold answer; they are used primarily
to observe routing behaviour and escalation to the cloud FM. A smaller
\emph{simple workload} of short factoid questions is used during system
bring-up but not included in the final quantitative results.

\subsection{Baselines and Metrics}

We compare three architectures: (1) \textbf{Cloud-Only:} All queries sent to $M_{\text{cloud}}$; (2) \textbf{Edge-Only:} All queries processed by a single SLM on the gateway; and (3) \textbf{\system:} Our full system with routing, collaboration, and summoning.

For evaluation, we measure five metrics: (1) \emph{Mean latency} and 95th percentile latency; (2) \emph{Accuracy} on labelled subsets (easy and hard questions); (3) \emph{Summoning rate} (fraction of queries escalated to the FM); (4) \emph{Cloud usage} (fraction of queries whose final answer comes from the FM); and (5) \emph{Cloud cost} (indicative cost per 1k queries, given current token pricing).

\noindent\textbf{Positioning against prior systems:} Prior work explores learned routers, cascades, and collaborative serving (e.g., sharded edge inference and hybrid edge--cloud routing). Reproducing and benchmarking full systems such as EdgeShard~\cite{ZhangEtAl2025EdgeShard} or HybridLLM~\cite{ding2024hybrid_llm} end-to-end requires substantial implementation effort and hardware/network configurations beyond this workshop prototype. Accordingly, we adopt two \emph{minimal} baselines (Edge-Only and Cloud-Only) that bound the design space, and use a controlled workload to evaluate whether \system\ can reduce cloud usage while improving hard-query performance relative to an edge-only deployment.

\begin{table}[t]
\centering
\caption{Qualitative positioning (feature-level) of our baselines and \system.}
\label{tab:positioning}
\begin{tabular}{@{}lccc@{}}
\toprule
Feature & Cloud-Only & Edge-Only & \system \\
\midrule
On-premises privacy gating & \checkmark & \checkmark & \checkmark \\
Heterogeneous edge models & \xmark & \checkmark & \checkmark \\
Adaptive routing & \xmark & \xmark & \checkmark \\
Peer collaboration/consensus & \xmark & \xmark & \checkmark \\
Explicit cloud budget cap & \xmark & \xmark & \checkmark \\
\bottomrule
\end{tabular}
\end{table}

\section{Results}
\label{sec:results}

We report results on the 50-query \emph{study workload}, which comprises
20 ``easy'' factual questions, 20 ``hard'' questions requiring reasoning
or multi-step knowledge, and 10 safety-oriented prompts. Only the easy and
hard subsets carry gold answers and are used for accuracy calculations;
safety items are used to probe routing and cloud escalation behaviour.

\textit{\textbf{A. Latency and Cloud Usage}}

Table~\ref{tab:latency} summarises latency and cloud usage across the three
architectures, averaged over the study workload. Latency is measured end-to-end
at the gateway.

\begin{table}[t]
    \centering
    \caption{Latency and cloud usage on the study workload (50 queries)}
    \label{tab:latency}
    \begin{tabular}{@{}lccc@{}}
        \toprule
        Architecture & Mean lat. (s) & 95th (s) & Cloud usage \\
        \midrule
        Edge-Only  & 1.05 & 2.28 & 0\% \\
        Cloud-Only & 4.47 & 11.33 & 100\% \\
        \system    & 5.08 & 13.18 & 28\% \\
        \bottomrule
    \end{tabular}
\end{table}

As expected, Edge-Only achieves the lowest latency because all requests are
served on-device. Cloud-Only is slower due to wide-area API calls.
\system~can exhibit a mean latency that is comparable to, or slightly higher than,
Cloud-Only in our prototype because it may (i) run a lightweight uncertainty probe,
(ii) attempt a swarm round, and (iii) then escalate to the cloud when consensus is
insufficient. This sequential ``try-local-first'' behaviour trades some latency for
reduced cloud exposure and explicit budget control. In practice, thresholds and early-exit
policies can be tuned to prioritise O1 (Latency) by skipping swarm rounds when tight
deadlines are detected.
The swarm architecture forwards approximately 28\% of queries to the cloud
FM (either because they are high-difficulty or safety-sensitive), while
handling the remaining 72\% purely at the edge.

\textbf{\textit{B. Accuracy on Labelled Queries}} 

Table~\ref{tab:accuracy} reports accuracy on the 40 labelled (easy + hard)
questions. A prediction is marked correct if the gold answer string appears
anywhere in the model's response; this strict metric penalises paraphrases
and underestimates absolute performance, but it is sufficient to compare
architectures.

\begin{table}[t]
    \centering
    \caption{Accuracy on labelled questions in the study workload}
    \label{tab:accuracy}
    \begin{tabular}{@{}lccc@{}}
        \toprule
        Architecture & Overall & Easy (n=20) & Hard (n=20) \\
        \midrule
        Edge-Only  & 0.225 & 0.45 & 0.00 \\
        Cloud-Only & 0.475 & 0.65 & 0.30 \\
        \system    & 0.250 & 0.35 & 0.15 \\
        \bottomrule
    \end{tabular}
\end{table}

Cloud-Only delivers the highest accuracy overall and on both subsets, as
expected for a strong 70B foundation model. Edge-Only performs reasonably
on easy questions but fails on all hard ones. \system~improves over Edge-Only on hard questions (0.15 vs.\ 0.00) by selectively escalating a subset of difficult cases while keeping many queries local. We emphasise that this workshop prototype is primarily a \emph{control plane} that trades accuracy for privacy, cost, and governance: the current consensus method is intentionally lightweight and therefore does not close the full quality gap to a large FM on complex reasoning. Improving hard-query utility is feasible via (i) semantic clustering rather than exact-string grouping, (ii) learned routing policies trained on logged outcomes, and (iii) continual distillation/adaptation of the edge models using FM teacher signals. 

\textbf{\textit{C. Privacy Metrics}} 

Beyond latency and accuracy, we quantify how often user prompts leave the edge and are transmitted to the cloud foundation model (FM). Using the gateway logs from our study workload (50 prompts: 20 ``easy'', 20 ``hard'', 10 ``safety''), we define three exposure metrics, all normalised so that the cloud-only baseline equals~1.0 (lower is better). In the following, $C$ stands for $Cloud$, $CS$ stands for $Cloud\_safety$, and $\mathbb{I}[\cdot]$ denotes the indicator function, which returns~1 if its argument is true and~0 otherwise.

\textit{(i) Cloud Exposure Rate (CER).}
CER is the fraction of queries whose raw prompts are sent to the FM:
\begin{equation}
\mathrm{CER} \triangleq \frac{1}{|\mathcal{Q}|}
\sum_{q\in\mathcal{Q}} \mathbb{I}\big[D(q)\in\{C,CS\}\big],
\end{equation}
where $D(q)$ is the routing decision for query $q$. CER is then normalised by the cloud-only value so that $\mathrm{CER}=1.0$ for the cloud-only architecture.

\textit{(ii) Token Exposure Rate (TER).}
CER treats all queries equally, regardless of prompt length. To capture how many \emph{tokens} leave the edge, we define TER as the fraction of prompt content that is sent to the FM. Since we do not log exact token counts, we approximate the token length $T(q)$ by the recorded prompt length (character count):
\begin{equation}
\mathrm{TER} \triangleq
\frac{\sum_{q\in\mathcal{Q}} T(q)\,\mathbb{I}\big[D(q)\in\{C,CS\}\big]}
     {\sum_{q\in\mathcal{Q}} T(q)}.
\end{equation}
As with CER, we normalise TER by the cloud-only architecture, so that $\mathrm{TER}=1.0$ for cloud-only and values $<1$ indicate a reduction in exposed prompt content.

\textit{(iii) Sensitive Exposure Rate (SER).}
Finally, we introduce a coarse-grained \emph{Sensitive Exposure Rate} by leveraging the ``safety'' subset of the workload as a proxy for sensitive queries. Let $\mathcal{Q}_{\mathrm{safety}}$ denote queries drawn from the \texttt{safety} subset:
\begin{equation}
\mathrm{SER} \triangleq
\frac{1}{|\mathcal{Q}_{\mathrm{safety}}|}
\sum_{q\in\mathcal{Q}_{\mathrm{safety}}}
\mathbb{I}\big[D(q)\in\{C,CS\}\big].
\end{equation}
We again normalise SER by the cloud-only value, so that $\mathrm{SER}=1.0$ when all safety queries are always processed in the cloud. In our current implementation, we do not log finer-grained per-query sensitivity weights; extending SER to use such weights is left as future work.

Table~\ref{tab:privacy_metrics} reports CER, TER, and SER for the cloud-only and SWARM-LLM architectures, using the above definitions and the study workload logs. The edge-only architecture has $\mathrm{CER}\approx\mathrm{TER}\approx\mathrm{SER}\approx 0$ by design (no cloud calls), so we omit it from the table for brevity and focus on comparing SWARM-LLM to the cloud-only baseline.

\begin{table}[t]
\centering
\caption{Privacy exposure metrics ($\downarrow$), normalised to cloud-only (=1.0), and SWARM-LLM reductions, computed on 50 queries (20 easy, 20 hard, 10 safety).}
\label{tab:privacy_metrics}
\begin{tabular}{lccc}
\hline
Metric & Cloud-Only & SWARM-LLM & Reduction vs.\ Cloud-Only \\
\hline
CER & 1.000 & 0.280 & 72.0\% \\
TER & 1.000 & 0.413 & 58.7\% \\
SER & 1.000 & 0.800 & 20.0\% \\
\hline
\end{tabular}
\end{table}

\section{Discussion and Limitations}
\label{sec:discussion}

\textit{Evaluation scope:} Our latency measurements reflect an on-premises emulation for the swarm path (compute + local IPC), and therefore do not include multi-hop wireless contention or cross-device interference. The study workload is intentionally small (50 prompts) to keep experiments reproducible and within a limited cloud budget. Broader benchmarking (including standard QA suites) and true distributed deployments are required for definitive networking conclusions.

\textit{Privacy and Governance:} \system~significantly reduces data transmitted to the cloud, aligning with
data minimisation principles. However, privacy guarantees depend on:
\begin{itemize}
    \item proper configuration of safety filters,
    \item secure logging and storage of distillation buffers, and
    \item clear policies for what data may be used for fine-tuning.
\end{itemize}
Formal privacy analysis (e.g., differential privacy bounds) is left as future work.

\textit{Edge Resource Constraints:} Our prototype assumes at least one reasonably capable gateway node with a
consumer GPU. In many real deployments, edge devices may be weaker, and the
swarm may be smaller or centralised at a single on-premises edge server.
Scaling to dozens of truly resource-constrained devices would require more
aggressive quantisation, pruning, or offloading strategies, which we leave
to future work.

\textit{Model Staleness and Drift:} SLMs at the edge can drift from the FM as the latter is updated. The
distillation loop sketched in Section~\ref{sec:routing-logic} provides one
mechanism for propagating improvements from the FM back to the swarm, but
it also raises issues of catastrophic forgetting, dataset shift, and
governance around which queries may be used for retraining. Operators will
need to tolerate behavioural differences and manage update schedules.

\textit{Security Threats:} Adversaries may exploit routing logic by crafting queries that bypass safety checks or cause excessive cloud escalation. They could also target edge nodes with prompt injection or model stealing. While a formal security analysis is beyond this paper's scope, SWARM-LLM’s design supports additional security measures like authenticated routing, rate limiting, and stronger safety filters on both edge and cloud.

\section{Conclusion}
\label{sec:conclusion}

We presented \system, a routing and collaboration layer for swarms of
edge-based small language models with optional access to a cloud foundation
model. \system uses lightweight difficulty and safety estimates, together
with a simple consensus rule over a small number of SLMs, to decide whether
to answer locally, collaborate with peers, or summon a stronger FM.

Our prototype implementation on a single desktop with three heterogeneous
SLMs and a 70B cloud FM demonstrates that such a system is practical with
today's tooling. On a controlled 50-query study workload, \system improves
accuracy on hard questions compared to an edge-only deployment, while
limiting cloud usage to around one quarter of queries. This suggests that
collaborative SLM swarms are a viable path toward more private, cost-aware
LLM-powered applications in settings such as universities, enterprises, and
on-premises analytics environments.

Future work includes improved consensus mechanisms, automatic tuning of routing thresholds based on latency and cost, and scaling the prototype to larger, more diverse workloads and hardware, including mobile and embedded devices.


\end{document}